\documentclass[a4paper,UKenglish,cleveref,autoref]{lipics-v2021}
\title{Hardness, Approximation, and Parameterized Algorithms for Asynchronous Border Minimization}
\titlerunning{Algorithms for Asynchronous Border Minimization}
\author{Giulian Buzatu}{Faculty of Mathematics and Computer Science, University of Bucharest, Romania}{buzatugiulian@yahoo.com}{}{}
\author{Adrian Micl\u au\c s}{Faculty of Mathematics and Computer Science, University of Bucharest, Romania}{adrian.miclaus1@gmail.com}{}{}
\author{Alexandru Popa}{Faculty of Mathematics and Computer Science, University of Bucharest, Romania}{alexandru.popa@fmi.unibuc.ro}{}{}
\authorrunning{G. Buzatu, A. Micl\u au\c s, and A. Popa}
\Copyright{Giulian Buzatu, Adrian Micl\u au\c s, and Alexandru Popa}

\hideLIPIcs
\EventEditors{}
\EventNoEds{1}
\EventLongTitle{38th Annual Symposium on Combinatorial Pattern Matching (CPM 2027)}
\EventShortTitle{CPM 2027}
\EventAcronym{CPM}
\EventYear{2027}
\EventDate{June 16--18, 2027}
\EventLocation{London, United Kingdom}
\EventLogo{}
\SeriesVolume{}
\ArticleNo{}
\usepackage{amsthm,amsmath,amssymb,mathtools}
\usepackage{booktabs,tabularx,array,microtype}
\usepackage{tikz}
\newcommand{\BMP}{\textsc{BMP}}
\newcommand{\PBMP}{\textsc{P-BMP}}

\newcommand{\lcs}{\operatorname{lcs}}
\newcommand{\OPT}{\operatorname{OPT}}
\newcommand{\BL}{\operatorname{BL}}

\newcommand{\poly}{\operatorname{poly}}

\newcommand{\gap}{\mathtt{-}}
\newtheorem{problem}{Problem}
\crefname{problem}{Problem}{Problems}

\nolinenumbers
\begin{document}
\maketitle
\begin{abstract}
In the Asynchronous Border Minimization Problem, a set of short DNA strings must be synthesized on a grid. At each step, the machine offers one character. Each cell may accept it or skip it. A border is counted when two neighboring cells make different choices. The goal is to minimize the sum of these borders. We consider both free placement, where the strings may be rearranged, and fixed placement, where their positions are given.

We show that free-placement border minimization is NP-complete on square grids over an alphabet of three symbols. In one dimension, the problem is APX-hard over a binary alphabet, even when all strings have equal length. It is also NP-complete for strings of length two over an unbounded alphabet, and we give an $8/7$-approximation for this case. For square grids, we give a randomized $O(n^{1/4}\log n)$-approximation, improving the previously known $O(n^{1/4}\log^2 n)$ bound. We also show that both free- and fixed-placement problems on rectangular grids are fixed-parameter tractable in the allowed border cost alone.
\end{abstract}

\section{Introduction}\label{sec:intro}
\subsection{Motivation}
DNA microarrays are small surfaces containing many short DNA strings,
called \emph{probes}~\cite{Pease1994,Hannenhalli2002}. One way to
manufacture them is to synthesize all probes in parallel using
light-directed chemistry~\cite{Fodor1991,Pease1994}. At each step, a
machine offers one nucleotide, while a mask exposes only the cells
that should receive it~\cite{Guenoche2004,KahngEtAl2004}. A boundary
between an exposed cell and a protected neighboring cell creates a
border in the mask. Light scattering near these borders may cause
synthesis errors~\cite{FeldmanPevzner1994}. This motivates minimizing
the total border length over all steps~\cite{Kasif2002,KahngEtAl2004,
KahngEtAl2006}.

We may need to choose both where each probe is placed and when its
characters are deposited~\cite{Hannenhalli2002,Li2018}. Placing
similar probes next to one another may reduce the cost, but the best
placement also depends on the deposition schedule~\cite{KahngEtAl2004,
Li2018}.

\subsection{Informal problem definition}
Suppose two neighboring cells must produce $AC$ and $CA$. The machine can offer $D=ACA$. The first cell accepts the first two characters and the second cell accepts the last two. We write their choices as $AC\gap$ and $\gap CA$, where $\gap$ means that the cell skips a step. The cells make different choices in the first and last steps, so their common boundary contributes $2$ to the cost.

The string $D$ is called the \emph{deposition sequence}. The choices made by one cell form an \emph{embedding} of its probe into $D$. A cell may skip a character even when it is the next character it needs; this is the unrestricted asynchronous model. In \BMP{}, we may choose the positions of the probes as well as the deposition sequence and embeddings. In \PBMP{}, their positions are fixed. We also study the corresponding one-dimensional problems, in which neighboring positions form a path.

\subsection{Previous and related work}
Early work studied algorithms for DNA array placement and mask
design~\cite{Hannenhalli2002}, as well as methods that combine probe
placement and embedding~\cite{CarvalhoRahmann2008}.
Li et al.~\cite{LiWXY2008} gave approximation algorithms for
asynchronous border minimization. Later, Li, Popa, Wong, and
Yung~\cite{Li2018} proved that two-dimensional \BMP{} and \PBMP{},
and free-placement one-dimensional \BMP{}, are NP-hard.
They also gave a polynomial-time algorithm for fixed-placement
one-dimensional \BMP{}, a randomized $O(n^{1/4}\log^2 n)$-approximation
for square-grid \BMP{}, a randomized $O(\log n)$-approximation
for \PBMP{}, and a $3/2$-approximation for one-dimensional \BMP{}.

Ganian et al.~\cite{Ganian2019} studied the parameterized complexity
of related problems with \emph{exhaustive} masks. In their model,
a cell must receive an offered character whenever it matches
the next required character. Our setting allows a cell to skip
such a character. Therefore, their algorithms do not directly
apply to the unrestricted problems studied here.

\subsection{Our results}
We first study hardness under restrictions on the alphabet and the length of the probes. A reduction from a binary Shortest Common Supersequence problem with prescribed numbers of zeros and ones shows that square-grid \BMP{} is NP-complete with three symbols. For one-dimensional \BMP{}, we prove APX-hardness over a binary alphabet and NP-completeness when every probe has length two.

We then give approximation algorithms. For square-grid \BMP{}, we improve the known ratio from $O(n^{1/4}\log^2 n)$ to $O(n^{1/4}\log n)$. The main idea is to choose one tree approximation of the string distances and use it both to place and to align the probes. We also obtain approximation guarantees for equal-length probes, fixed placements with few distinct strings, narrow grids, and length-two probes in one dimension.

Finally, we study exact algorithms. We show that \BMP{} and \PBMP{} on rectangular grids are fixed-parameter tractable in the border budget $b$, without bounding the alphabet or the probe lengths. We give polynomial-time algorithms for square-grid instances with two distinct strings and an exact algorithm for one-dimensional instances with few distinct strings. The definitions and notation used in these results are given next.

\section{Preliminaries}\label{sec:prelim}
Let $\Sigma$ be a finite alphabet and $S=(s_1,\ldots,s_n)$ a multiset of strings over $\Sigma$. We call each occurrence of a string a \emph{probe}. Equal strings may occur several times and are treated as different probes when placing them.

A \emph{deposition sequence} is a string $D=d_1\cdots d_T\in\Sigma^*$. At step $t$, the machine offers the character $d_t$ to all cells. An \emph{embedding} of probe $s_i$ into $D$ is a string $\varepsilon_i\in(\Sigma\cup\{\gap\})^T$ obtained by inserting gaps into $s_i$, with $\varepsilon_i[t]\in\{d_t,\gap\}$ for all $t\in[T]$. Removing the gaps gives exactly $s_i$. Thus $\varepsilon_i[t]=\gap$ means that probe $i$ receives nothing in step $t$.

Let $G=(V,E)$ be an undirected \emph{interaction graph}. Its vertices are cells, and an edge joins two cells whose common boundary is counted. Write $s_v$ and $\varepsilon_v$ for the probe and embedding assigned to vertex $v$. The cost on an edge $uv$ is $h_\varepsilon(u,v)=|\{t\in[T]:\varepsilon_u[t]\ne\varepsilon_v[t]\}|$, and the total \emph{border length} is $\BL_G(\varepsilon)=\sum_{uv\in E}h_\varepsilon(u,v)$. Only boundaries between adjacent cells count; the outer perimeter is excluded. For $A\subseteq V$, let $\delta_G(A)$ be the set of edges with exactly one endpoint in $A$. If precisely the cells in $A$ accept the offered character at one step, that step costs $|\delta_G(A)|$.

Let $G_{r,c}$ be the $r\times c$ grid on $[r]\times[c]$, with edges between horizontally or vertically consecutive cells. Write $G_m=G_{m,m}$ for a square grid.

\begin{problem}[Border Minimization Problem]\label{prob:bmp}
Given a multiset $S=(s_1,\ldots,s_n)$ of $n=rc$ probes and an $r\times c$ grid $G_{r,c}$, choose a bijection $\varphi:[n]\to V(G_{r,c})$, a deposition sequence $D$, and valid embeddings of the probes into $D$ so as to minimize $\BL_{G_{r,c}}$.
\end{problem}

\begin{problem}[Placement Border Minimization Problem]\label{prob:pbmp}
Given an $r\times c$ grid $G_{r,c}$ with a prescribed probe $s_v$ at every vertex $v$, choose a deposition sequence $D$ and valid embeddings so as to minimize $\BL_{G_{r,c}}$, without changing the placement.
\end{problem}

For example, consider $S=\{AC,AC,CA,CA\}$ on a $2\times2$ grid. Using $D=ACA$ and the embeddings $AC\gap$ and $\gap CA$, placing equal probes in the same row gives cost $4$, while placing them at opposite corners gives cost $8$. The first placement is available in \BMP{}, but not necessarily in \PBMP{}.

The one-dimensional versions use the path $P_n$. Only consecutive positions are adjacent.

\begin{problem}[One-Dimensional Border Minimization Problem]\label{prob:1dbmp}
Given a multiset $S=(s_1,\ldots,s_n)$ of probes, choose an order $\pi$ of the probes, a deposition sequence, and valid embeddings so as to minimize $\sum_{i=1}^{n-1}h_\varepsilon(\pi(i),\pi(i+1))$.
\end{problem}

\begin{problem}[One-Dimensional Placement Border Minimization Problem]\label{prob:1dpbmp}
Given an ordered sequence $(s_1,\ldots,s_n)$ of probes, choose a deposition sequence and valid embeddings so as to minimize $\sum_{i=1}^{n-1}h_\varepsilon(i,i+1)$, keeping the input order.
\end{problem}

For instance, $(AC,CA,AC)$ has minimum border cost $4$, whereas the freely chosen order $(AC,AC,CA)$ has cost $2$.

The decision version of each problem also receives a \emph{border budget} $b\ge0$ and asks whether the cost can be at most $b$. Throughout the paper, $n$ is the number of probes, $N$ is the total input size, $k$ is the number of distinct probe strings, $\ell$ is the maximum probe length, and $\OPT$ is the minimum border cost.

For strings $x,y$, let $\lcs(x,y)$ be the length of a \emph{longest common subsequence} (LCS). Define their \emph{insertion--deletion distance} by $d(x,y)=|x|+|y|-2\lcs(x,y)$. It is the minimum number of character insertions and deletions needed to turn one string into the other. For equal-length strings, $\operatorname{Ham}(x,y)$ is their Hamming distance, the number of positions in which they differ. Both distances can be computed in polynomial time. If $x,y$ are embedded in the same deposition sequence, their shared non-gap positions form a common subsequence, so their border cost is at least $d(x,y)$. In particular, every fixed placement satisfies $\OPT\ge\sum_{uv\in E}d(s_u,s_v)$.

In general, an optimal alignment for each neighboring pair may not fit into a single deposition sequence. Trees are an important exception.

\begin{lemma}\label{lem:treealign}
For every edge $uv$, any valid embeddings satisfy $h_\varepsilon(u,v)\ge d(s_u,s_v)$. For any tree $H$ on the probe occurrences, one can construct in polynomial time a common deposition sequence and embeddings such that $h_\varepsilon(u,v)=d(s_u,s_v)$ for every edge $uv$ of $H$. For any vertices $x,y$, these embeddings also satisfy $h_\varepsilon(x,y)\le d_H(x,y)$, where $d_H(x,y)$ is the sum of the $d$-distances along the unique path from $x$ to $y$ in $H$.
\end{lemma}

\begin{proof}
The first inequality follows from the definition of $d$. To prove the second claim, root $H$ and start with the root probe as the deposition sequence. Process each child after its parent. When adding a vertex $v$ with parent $u$, compute an LCS of $s_u$ and $s_v$. Match its characters to the corresponding entries of the already embedded probe $u$. Add deposition columns for the other characters of $v$ and place gaps in all earlier embeddings at those columns.

The embeddings of $u$ and $v$ share exactly $\lcs(s_u,s_v)$ non-gap entries. Their cost is therefore $d(s_u,s_v)$. Newly inserted columns do not change the costs of earlier tree edges. This realizes all tree-edge distances at once. Finally, the triangle inequality for Hamming distance between padded embeddings gives $h_\varepsilon(x,y)\le d_H(x,y)$.
\end{proof}

\section{Reductions and Hardness}\label{sec:hardness}
\subsection{BMP over a ternary alphabet}
\begin{problem}[Fixed-Count Binary Shortest Common Supersequence]\label{prob:source-scs}
Given binary strings $x_1,\ldots,x_k\in\{0,1\}^*$ and nonnegative integers $p,q$, decide whether there exists a binary string $z$ containing every $x_i$ as a subsequence and having exactly $p$ occurrences of $0$ and exactly $q$ occurrences of $1$.
\end{problem}

A \emph{supersequence} of $x$ is a string that contains $x$ as a subsequence. The counts $p,q$ refer to the numbers of symbols in the entire supersequence, not to the numbers used by any single input word. Requiring exactly $p$ zeros and $q$ ones is equivalent to requiring at most these counts, because unused symbols may be appended. The fixed-count binary restriction is NP-hard by the results of Middendorf~\cite{Middendorf1995}.

We establish hardness of the free-placement \BMP{} (\cref{prob:bmp}) by reducing from \cref{prob:source-scs}. The reduction has two stages. First, we turn prescribed supersequence counts into a fixed-placement \PBMP{} instance: two rows force the available numbers of zeros and ones, while a third row contains the input strings. Second, we reduce this fixed-placement instance to free-placement \BMP{}. Long coordinate prefixes make changing the intended placement too expensive, and a separator prevents those prefixes from being aligned with the original strings.
\begin{lemma}\label{lem:prefix}
For any fixed undirected interaction graph $G$ and any word $w$, replacing every vertex word $s_v$ by $ws_v$ leaves the optimum unrestricted fixed-placement border cost unchanged. The analogous assertion holds for common suffixes.
\end{lemma}
\begin{proof}[Proof sketch]
Prepending the same character to every probe can only add a column in which all cells are active. Conversely, removing the first occurrence of that character from each row cannot increase the cost on any edge. Repeating this argument proves the prefix claim; reversing the words proves the suffix claim. The full proof is in Appendix~\ref{app:lem-prefix}.
\end{proof}

\begin{lemma}\label{lem:binarypbmp}
There is a reduction from Fixed-Count Binary Shortest Common Supersequence (\cref{prob:source-scs}) to binary \PBMP{} (\cref{prob:pbmp}) with nonempty words. If the source instance has $k$ strings, total length $H$, and prescribed counts $p,q$, the resulting $(2k+1)\times(2k+1)$ instance has border optimum at least $B_0=2(2k+1)(p+q)+2H$, with equality if and only if the SCS instance is feasible.
\end{lemma}

\begin{proof}
We first ensure that the required counts are at most the total numbers of the corresponding symbols in the source strings. Indeed, in a feasible supersequence, retain only positions used by at least one input string. There are at most $H_0=\sum_i|x_i|_0$ retained zeros and $H_1=\sum_i|x_i|_1$ retained ones. Extra symbols can always be appended. Thus we may work with $p\le H_0$ and $q\le H_1$, which also ensures polynomially bounded counts. We may assume $k\ge2$ by repeating one input string.

Set $r=2k+1$ and use an $r\times r$ grid. Put $0^p$ in every cell of row 2, and $1^q$ in every cell of row 4. Row 3 contains $\epsilon,x_1,\epsilon,x_2,\ldots,\epsilon,x_k,\epsilon$. All other cells contain the empty string. We call rows 2 and 4 the \emph{anchor rows}. Their probes ensure that exactly $rp$ zeros and $rq$ ones are deposited in those rows.

Consider a step offering $0$. Let $R$ be the set of active column positions in row 2 and $T$ those in row 3. The positions in $T$ are separated by empty cells, so row 3 has $2|T|$ horizontal borders. Let $h(R)$ be the number of horizontal borders inside row 2. Counting the three boundaries between rows 1--4 and the horizontal borders gives $|R|+|R\triangle T|+3|T|+h(R)$, which equals $2|R|+2|T|+2|T\setminus R|+h(R)$.
The same calculation applies to a step offering $1$, using row 4 as the anchor. Let $H=\sum_i|x_i|$. Summing the expression over all deposition steps gives $\BL=B_0+2U+V$, with
$B_0=2r(p+q)+2H$, $U\ge0$ counting source deposits outside their corresponding active anchor positions, and $V\ge0$ counting horizontal borders in the anchor rows.

If $\BL=B_0$, then $U=V=0$. Since an active subset of a row has no horizontal border only when it is the whole row, every nonempty anchor mask is a full row. Moreover, each character of every $x_i$ is deposited in a step in which the matching anchor row is active. Reading the full-row anchor steps in order gives a common supersequence with exactly $p$ zeros and $q$ ones. Conversely, such a supersequence lets us deposit each character of $x_i$ during the matching anchor step, with no extra cost. Thus the border cost equals $B_0$ exactly when the SCS instance is feasible. Finally, add the same one-letter prefix to all probes. By \cref{lem:prefix} this makes them nonempty without changing the optimum.
\end{proof}

\begin{lemma}\label{lem:rigidity}
Let $G_m$ be the $m\times m$ grid, with $M=2m(m-1)$ edges. For a bijective relabeling $\pi$ of the coordinate pairs $[m]^2$, put $C(\pi)=\sum_{uv\in E(G_m)}\bigl\|\pi^{-1}(u)-\pi^{-1}(v)\bigr\|_1$. Then $C(\pi)\ge M$, with equality exactly when $\pi$ is a grid automorphism. Otherwise $C(\pi)\ge M+1$.
\end{lemma}

\begin{proof}[Proof sketch]
Each grid edge joins different coordinate labels and therefore contributes at least one. If equality holds, the relabeling preserves every grid edge, so it is a grid automorphism. The full proof is in Appendix~\ref{app:lem-rigidity}.
\end{proof}

\begin{lemma}\label{lem:forcing}
Given a fixed-layout binary \PBMP{} instance (\cref{prob:pbmp}) on an $m\times m$ grid, with maximum word length $L\ge1$, let $M=2m(m-1)$, $t=2ML+1$, $U_0=tM+2ML$ and $K=U_0+L+1$. Introduce a new letter $a$ and give the free-placement BMP instance the $m^2$ words $q_{ij}=a^{it}0^{jt}1^K s_{ij}$. Then $\OPT_{\BMP}(q)=tM+\OPT_{\PBMP}(s)$.
\end{lemma}

\begin{proof}
We first keep every input probe at its original position. Deposit all prefixes $a^{it}0^{jt}$, then deposit $1^K$ at every cell, and finally use an optimal deposition schedule for the original strings $s_{ij}$. The coordinate prefixes cost exactly $t$ on each grid edge, because its endpoints differ by one in one coordinate. Thus this construction has cost $tM+\OPT_{\PBMP}(s)$. Since one can always deposit the original characters separately, $\OPT_{\PBMP}(s)\le2ML$. The constructed solution therefore costs at most $U_0$.

We show that every solution of cost at most $U_0$ respects the intended separation between coordinate and payload characters. Suppose that one deposition step puts a \emph{coordinate zero} in one endpoint of a grid edge and a \emph{payload zero} in the other. The first endpoint has not yet deposited its block $1^K$, while the second has already finished that block. Of the first endpoint's $K$ separator ones, at most $L$ can be matched to payload ones at the second endpoint. The edge must therefore have cost at least $K-L>U_0$, which is impossible.

In any column containing both coordinate and payload zeros, their sets of active vertices consequently have no grid edge between them. We can split this column into two: one for coordinate zeros and one for payload zeros. The border cost does not change, since the cut sizes of these two sets add up to the cut size of their union. Repeating this operation separates the coordinate prefixes from the common separator and the payloads. No probe order is changed by the split.

Now consider an arbitrary placement $\pi$. On an edge joining probes labelled $(i,j)$ and $(i',j')$, the coordinate part costs at least $t(|i-i'|+|j-j'|)$: the counts of $a$ and of coordinate zeros differ by these amounts. Summing over all grid edges gives at least $tC(\pi)$. If $\pi$ is not a grid automorphism, \cref{lem:rigidity} gives $tC(\pi)\ge t(M+1)>U_0$. Thus every solution of cost at most $U_0$ uses a grid automorphism.

Such an automorphism preserves all original adjacencies. After the coordinate part is removed, the suffixes are $1^Ks_{ij}$. Their optimum cost is $\OPT_{\PBMP}(s)$ by common-prefix cancellation (\cref{lem:prefix}). Hence every solution costs at least $tM+\OPT_{\PBMP}(s)$, matching the first construction.
\end{proof}

\begin{theorem}\label{thm:ternary}
The decision version of free-placement \BMP{} (\cref{prob:bmp}) is NP-complete over an alphabet of three symbols, even for nonempty input probes.
\end{theorem}
\begin{proof}
Reduce an instance of Fixed-Count Binary Shortest Common Supersequence (\cref{prob:source-scs}) using Lemma~\ref{lem:binarypbmp}, obtaining one binary P-BMP instance $I$ and threshold $B_0$. Apply Lemma~\ref{lem:forcing} to $I$, obtaining one ternary BMP instance and threshold $tM+B_0$. By the exact optimum identities, the original source is feasible if and only if the single output BMP instance has border cost at most this threshold. The output length is polynomial: after count normalization $p+q\le H$, the P-BMP grid has side $O(k)$ and polynomial maximum string length; the coordinates and separator have lengths polynomial in those quantities. For NP membership, an arbitrary feasible synthesis can be shortened by removing empty deposition columns, leaving at most the total number of probe characters. Such a sequence of masks is a polynomial-size verifiable certificate.
\end{proof}

\subsection{One-dimensional hardness}
On a path, the tree-alignment lemma gives an exact characterization: for any fixed probe order $\pi$, the best possible border cost is $\sum_{i=1}^{n-1}d(s_{\pi(i)},s_{\pi(i+1)})$. Therefore 1D-\BMP{} is a shortest Hamiltonian \emph{path} problem in the insertion--deletion metric: visit every distinct probe exactly once, choosing an order that minimizes the sum of distances between successive strings. We use this property to prove hardness under two restrictions: binary strings of arbitrary length and strings of length two over an arbitrary alphabet.

\begin{problem}[Bounded-Degree Gap $(1,2)$-TSP]\label{prob:source-tsp}
Given an undirected graph $Q=(W,F)$ of maximum degree at most six, assign distance $1$ to each edge of $F$ and distance $2$ to every other pair of distinct vertices. For a fixed universal constant $\delta>0$, distinguish the following promised cases: $Q$ has a Hamiltonian cycle, or every cyclic ordering of its $m=|W|$ vertices contains at least $\delta m$ consecutive nonedges (and thus has tour cost at least $(1+\delta)m$).
\end{problem}

This is a promise problem: the input is guaranteed to belong to one of the two cases, and the algorithm must distinguish them. The distance of a tour is the sum of its consecutive pair distances, including the edge returning to its start. The required constant-gap hardness is recorded in the literature on $(1,2)$-TSP~\cite{PapadimitriouYannakakis1993,VanEeSitters2020}.

\begin{problem}[Edge Hamiltonian Path on Bipartite Graphs]\label{prob:source-ehp}
Given an undirected bipartite graph $Q=(U\cup W,F)$, decide whether its edges can be ordered as $e_1,\ldots,e_{|F|}$ so that every two consecutive edges $e_i,e_{i+1}$ share an endpoint.
\end{problem}

In an edge Hamiltonian path, edges need not form a vertex-simple path; only consecutive edges must meet. This problem is NP-complete even for bipartite graphs~\cite{LaiWei1993}.

\begin{lemma}\label{lem:binary}
For $x\in\{0,1\}^d$, put $T=4d+3$, $B(0)=1 0^T 11$, $B(1)=11 0^T 1$, and $F(x)=B(x_1)0^T B(x_2)\cdots 0^T B(x_d)$. For all $x,y\in\{0,1\}^d$ the words $F(x),F(y)$ have equal length and $d(F(x),F(y))=2\operatorname{Ham}(x,y)$.
\end{lemma}
\begin{proof}[Proof sketch]
The zero separators are long enough that every longest common subsequence must align the corresponding blocks. Matching the short runs then loses exactly one common character for each differing bit, giving the stated identity. The full proof is in Appendix~\ref{app:lem-binary}.
\end{proof}

\begin{theorem}\label{thm:binaryapx}
There exists an absolute $\varepsilon>0$ such that approximating free-placement 1D-\BMP{} (\cref{prob:1dbmp}) over a binary alphabet within $1+\varepsilon$ is NP-hard, even when every probe has the same length.
\end{theorem}
\begin{proof}
We reduce from Bounded-Degree Gap $(1,2)$-TSP (\cref{prob:source-tsp}). Let $Q=(W,F)$ be its graph, with $n=|W|$ and maximum degree at most six. For each vertex $v$, make a binary vector $x_v$. It has one coordinate for every edge of $Q$ and, in addition, private coordinates used only by $v$. Set the coordinates of the edges incident with $v$ to one, then add $6-\deg_Q(v)$ private ones. Every vector now has exactly six ones.

If $uv\in F$, then $x_u$ and $x_v$ share exactly one coordinate equal to one, corresponding to their common edge. Otherwise they share none. Thus their Hamming distance is $10$ for an edge of $Q$ and $12$ for a nonedge. Apply the encoding of \cref{lem:binary} and put $s_v=F(x_v)$. The resulting probes are binary and have equal length, and their insertion--deletion distances are $20$ or $24$, respectively.

For any order $\pi$ of the probes, let $b(\pi)$ count consecutive pairs that are nonedges of $Q$. By \cref{lem:treealign}, the exact one-dimensional border cost of this order is $20(n-1)+4b(\pi)$.
If $Q$ has a Hamiltonian cycle, it has a Hamiltonian path, giving $b(\pi)=0$ and optimum $20(n-1)$. In the other promised case, closing any ordering into a cycle adds at most one nonedge. Hence $b(\pi)\ge\delta n-1$ for every order. For $n\ge2/\delta$, every solution then costs at least $20(n-1)+2\delta n\ge(1+\delta/10)20(n-1)$.

This is a constant gap between the two cases. The finitely many smaller instances may be decided directly. Therefore a polynomial-time approximation with ratio smaller than $1+\delta/10$ would distinguish the promised cases. This proves APX-hardness. The existing $3/2$-approximation of Li et al.~\cite{Li2018} also places the one-dimensional problem in APX.
\end{proof}

\begin{theorem}\label{thm:length2hard}
Free-placement 1D-\BMP{} (\cref{prob:1dbmp}) is NP-complete even when every probe has length exactly two, if the alphabet is unbounded.
\end{theorem}
\begin{proof}
We reduce from Edge Hamiltonian Path on Bipartite Graphs (\cref{prob:source-ehp}), shown NP-complete by Lai and Wei~\cite{LaiWei1993}. For $G=(U\cup V,E)$, make the alphabet $U\cup V$ and a length-two probe $s_{uv}=uv$ for every edge $uv\in E$ (with $u\in U$). Distinct edge probes have LCS length one if and only if their graph edges share an endpoint, and otherwise LCS length zero. Thus their insertion/deletion distance is two or four, respectively. A 1D order has cost $2(|E|-1)+2h$, where $h$ is the number of successive nonincident edge pairs. The cost threshold $2(|E|-1)$ is achieved precisely if the original graph has an Edge Hamiltonian Path. NP membership follows from the polynomial certificate.
\end{proof}

\section{Approximation Algorithms}\label{sec:approx}
\subsection{Square-grid BMP}
For any placement $\varphi$, define its \emph{metric placement cost} $P_d(\varphi)=\sum_{uv\in E(G_m)}d(s_{\varphi^{-1}(u)},s_{\varphi^{-1}(v)})$. This is a lower bound on the border cost of every embedding using that placement. We approximate the metric $d$ by a random tree metric $d_T$. A \emph{$2$-hierarchically separated tree} (2-HST) groups the probes into nested clusters, with cluster scales falling by a factor of at least two at each level. Two probes have tree distance determined by the smallest cluster containing both. The sampled tree dominates $d$, but its expected distances grow by only an $O(\log n)$ factor~\cite{FRT2004}. We use this \emph{same tree} to determine both the grid placement and the alignment, so that only one logarithmic factor is lost.
\begin{lemma}\label{lem:hst}
Let $T$ be a rooted $2$-HST on the probe labels, with leaf-to-leaf ultrametric $d_T$ dominating their insertion--deletion distance $d$. There is a computable spanning tree $H$ on the original labels, with edge lengths $d$, such that $d_H(x,y)\le4d_T(x,y)$ for all leaves $x,y$. An absolute constant in place of $4$ also suffices if another standard $2$-HST normalization is used.
\end{lemma}
\begin{proof}
For every cluster $C$ of $T$, choose a representative $r(C)$ that is one of the leaves of $C$. Choose it recursively from one designated child of $C$. For each other child cluster $B$, add the edge $r(C)r(B)$, with weight $d(r(C),r(B))$. The edges inside the children, together with these new edges, form a spanning tree $H$ on the original probe labels.

Let $\Delta(C)$ be the distance in $d_T$ between leaves whose lowest common ancestor is $C$. Since $d_T$ dominates $d$, each new edge added within $C$ has $d$-weight at most $\Delta(C)$. The representative path from a leaf to a higher cluster uses at most one edge per cluster level. These scales decrease by a factor of at least two as we move down the hierarchy.

Let $C$ be the lowest common ancestor of leaves $x,y$. Their path in $H$ consists of a path from $x$ to $r(C)$ and one from $r(C)$ to $y$. Each has total weight at most $\sum_{j\ge0}\Delta(C)2^{-j}=2\Delta(C)$. Therefore $d_H(x,y)\le4\Delta(C)=4d_T(x,y)$. Equal strings, which have distance zero, may first be merged and then restored as zero-weight leaves.
\end{proof}

\begin{lemma}\label{lem:euler}
Let $n=m^2$ probes be leaves of a weighted tree metric $d_T$. Listing the leaves in depth-first Euler order and placing them row by row in an $m\times m$ grid gives a layout $\phi_T$ with $P_{d_T}(\phi_T)\le O(n^{1/4})\min_\phi P_{d_T}(\phi)$.
\end{lemma}
\begin{proof}[Proof sketch]
For a tree-edge cut whose smaller part has $s$ leaves, every grid layout crosses $\Omega(\sqrt{s})$ edges. A depth-first layout crosses $O(\min\{m,s\})$ edges. The ratio is $O(\sqrt m)=O(n^{1/4})$, and summing over the tree edges proves the bound. The full proof is in Appendix~\ref{app:lem-euler}.
\end{proof}

\begin{theorem}\label{thm:mainapprox}
There is a randomized polynomial-time $O(n^{1/4}\log n)$-approximation algorithm for square-grid \BMP{} (\cref{prob:bmp}) on $n$ probes of arbitrary length over an arbitrary alphabet.
\end{theorem}
\begin{proof}
Compute all pairwise insertion--deletion distances $d$. Draw a dominating $2$-HST $T$ from the Fakcharoenphol--Rao--Talwar distribution~\cite{FRT2004}. It satisfies $d(x,y)\le d_T(x,y)$ and $\mathbb E[d_T(x,y)]\le O(\log n)d(x,y)$ for each pair. Obtain the Euler layout $\phi_T$ of Lemma~\ref{lem:euler}. Convert $T$ to an actual probe spanning tree $H$ using Lemma~\ref{lem:hst}, and construct a single valid global probe alignment attaining every $H$-edge's $d$-distance using Lemma~\ref{lem:treealign}. Hamming triangle inequality on the aligned rows implies that the mismatch between two arbitrary probes is at most $d_H(x,y)\le4d_T(x,y)$. Consequently the total border cost of the constructed solution is at most $4P_{d_T}(\phi_T)$.

Fix the placement $\phi^*$ from an optimal BMP solution (independent of the random tree). Lemma~\ref{lem:euler} applies to each tree realization, so $\mathbb E[\BL]\le O(n^{1/4})\mathbb E[P_{d_T}(\phi^*)]\le O(n^{1/4}\log n)P_d(\phi^*)\le O(n^{1/4}\log n)\OPT_{\BMP}$. Pairwise LCS, FRT sampling, representative selection, and progressive tree alignment are polynomial-time procedures. This proves the asserted expected approximation bound.
\end{proof}

For probes of equal length $\ell$, an $O(\ell n^{1/4})$ approximation is also possible (see Proposition~\ref{prop:shortsync} in the appendix).

\subsection{Few probe types and short probes}
The approximation ratio for a fixed placement can depend on the number $k$ of \emph{different} probe strings rather than on the total number of cells. The algorithm aligns one copy of each type, then gives all copies of that type the same padded row. This is always a valid global deposition schedule, even when the original grid contains many copies.
The following algorithm also works if the fixed placement is represented by a more general interaction graph, not necessarily a grid. We merge equal words only while constructing their common alignment; the original number of grid adjacencies is preserved in the cost calculation.
\begin{theorem}\label{thm:fewtypes}
Let $k$ be the number of distinct probe words in a fixed-placement \PBMP{} instance (\cref{prob:pbmp}). There is a randomized polynomial-time $O(\log k)$ approximation for $k\ge2$. The same problem is exactly solvable when $k\le2$.
\end{theorem}
\begin{proof}[Proof sketch]
Choose one alignment for each distinct probe type and use it at every cell containing that type. A random tree approximation of the insertion--deletion distances, together with \cref{lem:hst,lem:treealign}, bounds the expected border cost by $O(\log k)\OPT$. With at most two types, a single LCS alignment attains the lower bound on every edge. Details are in Appendix~\ref{app:thm-fewtypes}.
\end{proof}

\begin{theorem}\label{thm:threetypes}
\PBMP{} (\cref{prob:pbmp}) with at most three distinct probe words admits a deterministic polynomial-time $4/3$ approximation on any fixed interaction graph, independent of the number of occurrences.
\end{theorem}
\begin{proof}[Proof sketch]
We try the three spanning trees on the three probe types and align each tree exactly. The triangle inequality bounds the additional cost on the omitted pair. The least expensive of the three solutions has cost at most $4\OPT/3$. The full proof is in Appendix~\ref{app:thm-threetypes}.
\end{proof}

\begin{theorem}\label{thm:eightseven}
If every input probe in 1D-\BMP{} (\cref{prob:1dbmp}) has length exactly two, there is a polynomial-time $8/7$ approximation even for an unbounded alphabet.
\end{theorem}
\begin{proof}[Proof sketch]
For length-two probes, the distance between two distinct strings is either $2$ or $4$. Make a graph on the distinct probe types, joining types at distance $2$. A maximum path cover in this graph gives an ordering with many distance-$2$ transitions. Using the known path-cover approximation and an exact alignment of the resulting order gives the ratio $8/7$. The full proof is in Appendix~\ref{app:thm-eightseven}.
\end{proof}

For a fixed-placement $r\times m$ grid, we also obtain a deterministic $(r+1)$ approximation. The construction and proof of Theorem~\ref{thm:thin} are in the appendix.

\section{Parameterized Algorithms}\label{sec:param}
\subsection{The border budget}
We use the budget $b$ to bound the number of relevant cells, even when the grid and probe lengths are large. First we solve alignments on a bounded number of cells. We then show how low-cost grids can be reduced to this case by removing repeated rows and columns.
\begin{lemma}\label{lem:banded}
For a connected fixed interaction graph on $q$ probes of maximum length $L$, a solution of border cost at most $b$ can be found, if one exists, in $(L+1)(2b+1)^{q-1}2^q\poly(q,L)$ time.
\end{lemma}
\begin{proof}
A standard exact global-alignment state records the vector $p=(p_v)_{v\in V}$ of consumed prefix lengths, starting at $0^q$. A legal transition chooses a character $c$ and a nonempty subset $A$ of vertices whose next undepleted character is $c$; it increments their positions and pays $|\delta_G(A)|$. Thus each directed transition increases the total number of consumed characters, so the state graph is acyclic. The minimum border cost is the shortest-path distance from the zero vector to $(|s_v|)_v$.

For every edge $uv$ and every reached state, the absolute consumption difference $|p_u-p_v|$ is at most the number of preceding columns in which exactly one of $u,v$ was active. This is the border cost accrued on edge $uv$, and is at most the global accumulated cost of a partial solution of budget $b$. Choose any root $r$ and any graph path from $r$ to $v$. Triangle inequality along that path gives $|p_v-p_r|\le b$ (since the distinct edge contributions sum to no more than the global budget). Consequently only vectors with $p_r\in[0,L]$ and $p_v\in[p_r-b,p_r+b]$ need be explored. They number at most $(L+1)(2b+1)^{q-1}$. At each state enumerate at most $2^q-1$ nonempty active subsets and check their next letters, retaining only transitions not exceeding budget $b$. This enumerates exactly every feasible cost-$b$ schedule because no such schedule ever leaves the band. The claimed bound follows.
\end{proof}

\begin{lemma}\label{lem:typecount}
Every YES-instance of free-placement BMP with budget $b$ on a connected interaction graph has at most $b+1$ distinct probe words.
\end{lemma}
\begin{proof}
Fix a solution and form the quotient graph of its distinct input-word types, connecting two types if occurrences lie on adjacent grid cells. The quotient graph is connected, and a spanning tree contains at least $k-1$ type-changing edges. Such an edge joins distinct words and therefore has insertion--deletion lower bound at least one. Thus $b\ge\OPT\ge k-1$.
\end{proof}

\begin{lemma}\label{lem:concentration}
In any border-cost-$b$ alignment on an $r\times c$ grid, at most $b$ rows contain a positive horizontal mismatch and at most $b$ columns contain a positive vertical mismatch. If $r,c>b$, all cells except at most $b^2$ carry one common \emph{padded} probe row.
\end{lemma}

\begin{proof}[Proof sketch]
At most $b$ rows can contain a positive horizontal mismatch, and at most $b$ columns a positive vertical mismatch. All other rows and columns have constant padded embeddings, and their intersections force these embeddings to agree. Only cells at intersections of exceptional rows and columns may differ. See Appendix~\ref{app:lem-concentration}.
\end{proof}

\begin{theorem}\label{thm:rectbmp}
For every $r\times c$ rectangle, the decision version of \BMP{} (\cref{prob:bmp}) with border budget $b$ is solvable in $2^{O(b^3\log(b+2))}\poly(N)$ time, where $N$ is the input size.
\end{theorem}
\begin{proof}
Reject if $k>b+1$ by \cref{lem:typecount}. We distinguish two cases.

\emph{Both dimensions exceed $b$.} By \cref{lem:concentration}, all but at most $b^2$ cells have one padded embedding $E^*$ and input word $s^*$. If the rectangle has more than $2b^2$ cells, $s^*$ is the unique strict-majority type; reject if no such type exists or there are more than $b^2$ non-$s^*$ occurrences. Replace the rectangle by one of dimensions $r'=\min\{r,2b+1\}$ and $c'=\min\{c,2b+1\}$, retaining all minority occurrences and filling other cells with $s^*$. If the original rectangle was already this small, do nothing.

For equivalence, call a row or column \emph{good} if all its padded entries equal $E^*$; there are at most $b$ bad rows and $b$ bad columns. Compress every maximal run of good rows or columns to one representative. No minority probe is lost, and the border cannot increase by the Hamming triangle inequality. At most $2b+1$ rows and columns remain; duplicate good rows or columns to obtain exactly $r'\times c'$, at zero cost. Conversely, any cost-$b$ embedding of the reduced instance has a common padded majority entry because $r',c'>b$ and, whenever compression occurred, $r'c'>2b^2$. It must represent $s^*$, the unique strict-majority type. Duplicate its good rows and columns to recover the original dimensions and multiplicities without additional border.

\emph{One dimension at most $b$.} Transpose if needed so $r\le b$. An ordered \emph{column pattern} records the input probe type in each of the $r$ rows of one column; it does not record an embedding. Each transition between different patterns costs at least one, so there are at most $b+1$ maximal runs. Each heterogeneous column has a positive vertical border, hence at most $b$ such columns occur. Truncate a homogeneous run longer than $b+1$ to exactly $b+1$ columns; deleting repeated-pattern columns never raises border by the Hamming triangle inequality. Conversely, among the retained $b+1$ columns some column has zero vertical border, so its padded entries coincide and it may be duplicated without extra border.

Enumerate all sequences of at most $b+1$ patterns over $k\le b+1$ types with retained run lengths at most $b+1$ and flags indicating truncated homogeneous runs. For each type $t$, let $a_t$ be its count in the bounded template and $m_t$ its original multiplicity. Require $m_t\ge a_t$, divisibility of $m_t-a_t$ by $r$, and an expandable retained run of type $t$ whenever this difference is positive. Such templates expand to the original multiset by inserting $(m_t-a_t)/r$ columns of type $t$. Every cost-$b$ original layout compresses to an enumerated template satisfying these conditions, and every budget-feasible template expands to a budget-feasible original layout.

There are $2^{O(b^2\log(b+2))}$ templates, each with $O(b^3)$ cells. Solve each using \cref{lem:banded}, obtaining $2^{O(b^3\log(b+2))}\poly(N)$ total time.
\end{proof}

\begin{theorem}\label{thm:rectpbmp}
\PBMP{} (\cref{prob:pbmp}) with a prescribed $r\times c$ layout is solvable in $2^{O(b^4\log(b+2))}\poly(N)$ time when parameterized only by target border cost $b$.
\end{theorem}
\begin{proof}
A \emph{row pattern} is the ordered list of probe strings in one row; it records the input strings, not their embeddings. If two consecutive rows have different patterns, some vertical edge between them joins different strings and has positive cost in every alignment. Hence an instance of cost at most $b$ has at most $b+1$ maximal runs of equal row patterns.

Call a row pattern \emph{homogeneous} if every probe in the row is the same string, and \emph{heterogeneous} otherwise. Every heterogeneous row has a positive horizontal border. Consequently, a YES-instance has at most $b$ occurrences of heterogeneous rows in total. In particular, a long run must be homogeneous.

Suppose a homogeneous run contains more than $b+1$ rows. Keep $b+1$ rows and delete the others. This cannot increase the cost of a given alignment: horizontal contributions are removed, and any new vertical boundary is bounded by the sum of the boundaries that it replaces. Conversely, in every cost-$b$ alignment of the reduced grid, at least one of these $b+1$ rows has zero horizontal border, since at most $b$ rows can have positive horizontal border. All padded embeddings on that row are identical. We may copy this row as often as needed inside the run without increasing the cost, restoring the original fixed placement.

Thus, if there are more than $b$ row-pattern changes or more than $b$ heterogeneous rows, we reject. Otherwise we may keep at most $(b+1)^2$ rows. Apply the same argument to the columns of the resulting grid. The reduced instance has at most $(b+1)^4$ cells, and every step can be reversed by duplicating zero-border rows or columns. We test it with \cref{lem:banded}. The running time is $2^{O(b^4\log(b+2))}\poly(N)$. This bounds the number of cells, not the total number of bits in the strings.
\end{proof}

\subsection{Few distinct strings in one dimension}
For 1D-\BMP{}, repetitions simplify the placement problem: all occurrences of a distinct probe string can be placed together at no cost between copies. We therefore solve a much smaller problem on the $k$ distinct strings, using a standard subset dynamic program for a shortest Hamiltonian path.
\begin{theorem}\label{thm:1dk}
For $k$ distinct input words and maximum probe length $L$, 1D-\BMP{} (\cref{prob:1dbmp}) has an exact algorithm in $O(nL+k^2L^2+2^kk^2)$ time, hence is FPT parameterized by $k$.
\end{theorem}
\begin{proof}[Proof sketch]
By the triangle inequality, copies of the same string can be placed consecutively in an optimal ordering. Compute the pairwise insertion--deletion distances of the $k$ distinct strings, then find a shortest Hamiltonian path by a subset dynamic program. An exact tree alignment realizes its cost. The full proof is in Appendix~\ref{app:thm-1dk}.
\end{proof}

Since a cost-$b$ YES-instance has at most $b+1$ distinct strings by \cref{lem:typecount}, the algorithm of \cref{thm:1dk} also establishes fixed-parameter tractability of 1D-\BMP{} in the border budget alone.

For free-placement square-grid \BMP{} with two distinct probe strings, an optimal solution can be found in polynomial time. The exact statement and proof are given as Theorem~\ref{thm:twotypes} in the appendix.

\section{Open Problems}\label{sec:conclusion}

Several questions about the complexity and approximability of asynchronous border minimization remain open. Does square-grid \BMP{} remain NP-hard when the alphabet is binary? Are \BMP{} and \PBMP{} APX-hard on square grids? On the algorithmic side, does square-grid \PBMP{} admit a constant-factor approximation independent of the number of probes? For square-grid \BMP{}, can the approximation ratio of $O(n^{1/4}\log n)$ be improved to $O(n^{1/4})$, or even to a constant? Finally, is square-grid \BMP{} fixed-parameter tractable when parameterized solely by the number of distinct probe strings?

\section*{AI disclosure} The authors used GPT-5.6 Sol during the preparation of this
work to brainstorm and refine constructions, explore proof
strategies, and discuss and check mathematical arguments.
The model also assisted with proofreading and aspects of
the literature review. All mathematical claims, proofs,
and cited results were independently verified by the authors,
who take full responsibility for the accuracy and content
of the paper.

\bibliography{references}

@article{Li2018,
  author={Cindy Y. Li and Alexandru Popa and Prudence W. H. Wong and Fencol C. C. Yung},
  title={Hardness and approximation of the asynchronous border minimization problem},
  journal={Discrete Applied Mathematics}, volume={235}, pages={101--117}, year={2018},
  doi={10.1016/j.dam.2017.08.025}
}

@article{Ganian2019,
  author={Robert Ganian and Martin Kronegger and Andreas Pfandler and Alexandru Popa},
  title={Parameterized Complexity of Asynchronous Border Minimization},
  journal={Algorithmica}, volume={81}, number={1}, pages={201--223}, year={2019},
  doi={10.1007/s00453-018-0442-5}
}

@article{FRT2004,
  author={Jittat Fakcharoenphol and Satish Rao and Kunal Talwar},
  title={A tight bound on approximating arbitrary metrics by tree metrics},
  journal={Journal of Computer and System Sciences}, volume={69}, number={3},
  pages={485--497}, year={2004}, doi={10.1016/j.jcss.2004.04.011}
}

@article{Middendorf1995,
  author={Martin Middendorf}, title={On finding minimal, maximal, and consistent sequences over a binary alphabet},
  journal={Theoretical Computer Science}, volume={145}, number={1--2}, pages={317--327},
  year={1995}, doi={10.1016/0304-3975(95)00014-N}
}

@article{PapadimitriouYannakakis1993,
  author={Christos H. Papadimitriou and Mihalis Yannakakis},
  title={The traveling salesman problem with distances one and two},
  journal={Mathematics of Operations Research}, volume={18}, number={1}, pages={1--11},
  year={1993}, doi={10.1287/moor.18.1.1}
}

@article{VanEeSitters2020,
  author={Martijn van Ee and Ren{\'e} Sitters}, title={The {Chinese} deliveryman problem},
  journal={4OR}, volume={18}, pages={341--356}, year={2020}, doi={10.1007/s10288-019-00420-2}
}

@article{LaiWei1993,
  author={Ten-Hwang Lai and Shu-Shang Wei},
  title={The edge {Hamiltonian} path problem is {NP}-complete for bipartite graphs},
  journal={Information Processing Letters}, volume={46}, number={1}, pages={21--26},
  year={1993}, doi={10.1016/0020-0190(93)90191-B}
}

@inproceedings{BermanKarpinski2006,
  author={Piotr Berman and Marek Karpinski}, title={$8/7$-Approximation Algorithm for $(1,2)$-{TSP}},
  booktitle={Proceedings of the Seventeenth Annual ACM--SIAM Symposium on Discrete Algorithms (SODA)},
  pages={641--648}, year={2006}, doi={10.1145/1109557.1109627}
}

@article{KahngEtAl2004,
  author={Andrew B. Kahng and Ion I. Mandoiu and Pavel A. Pevzner and Soheil Reda and Alexander Zelikovsky},
  title={Scalable heuristics for design of {DNA} probe arrays},
  journal={Journal of Computational Biology}, volume={11}, number={2--3},pages={429--447}, year={2004}
}

@article{KahngEtAl2006,
  author={Andrew B. Kahng and Ion I. Mandoiu and Soheil Reda and Xu Xu and Alexander Zelikovsky},
  title={Computer-aided optimization of {DNA} array design and manufacturing},
  journal={IEEE Transactions on Computer-Aided Design of Integrated Circuits and Systems},
  volume={25}, number={2}, pages={305--320}, year={2006}
}

@article{Kasif2002,
  author={Simon Kasif and Zhiping Weng and Antal Derti and Richard Beigel and Charles DeLisi},
  title={A computational framework for optimal masking in the synthesis of oligonucleotide microarrays},
  journal={Nucleic Acids Research}, volume={30}, number={20}, pages={e106}, year={2002},
  doi={10.1093/nar/gnf106}
}

@article{Fodor1991,
  author  = {Fodor, Stephen P. A. and Read, Jill L. and
             Pirrung, Michael C. and Stryer, Lubert and
             Lu, Amy T. and Solas, Dennis},
  title   = {Light-Directed, Spatially Addressable Parallel
             Chemical Synthesis},
  journal = {Science},
  volume  = {251},
  number  = {4995},
  pages   = {767--773},
  year    = {1991},
  doi     = {10.1126/science.1990438}
}

@article{Pease1994,
  author  = {Pease, Ann C. and Solas, Dennis and
             Sullivan, Elizabeth J. and Cronin, Michael T. and
             Holmes, Christopher P. and Fodor, Stephen P. A.},
  title   = {Light-Generated Oligonucleotide Arrays for Rapid
             {DNA} Sequence Analysis},
  journal = {Proceedings of the National Academy of Sciences},
  volume  = {91},
  number  = {11},
  pages   = {5022--5026},
  year    = {1994},
  doi     = {10.1073/pnas.91.11.5022}
}

@article{FeldmanPevzner1994,
  author  = {Feldman, W. M. and Pevzner, Pavel A.},
  title   = {Gray Code Masks for Sequencing by Hybridization},
  journal = {Genomics},
  volume  = {23},
  number  = {1},
  pages   = {233--235},
  year    = {1994},
  doi     = {10.1006/geno.1994.1482}
}

@article{Hannenhalli2002,
  author  = {Hannenhalli, Sridhar and Hubell, Earl and
             Lipshutz, Robert and Pevzner, Pavel A.},
  title   = {Combinatorial Algorithms for Design of {DNA} Arrays},
  journal = {Advances in Biochemical Engineering/Biotechnology},
  volume  = {77},
  pages   = {1--19},
  year    = {2002},
  doi     = {10.1007/3-540-45713-5_1}
}

@article{Guenoche2004,
  author  = {Gu{\'e}noche, Alain},
  title   = {Supersequences of Masks for Oligo-Chips},
  journal = {Journal of Bioinformatics and Computational Biology},
  volume  = {2},
  number  = {3},
  pages   = {459--469},
  year    = {2004},
  doi     = {10.1142/S0219720004000685}
}

@article{CarvalhoRahmann2008,
  author  = {S{\'e}rgio A. de Carvalho Jr. and Sven Rahmann},
  title   = {Better {GeneChip} Microarray Layouts by Combining
             Probe Placement and Embedding},
  journal = {Journal of Bioinformatics and Computational Biology},
  volume  = {6},
  number  = {3},
  pages   = {623--641},
  year    = {2008},
  doi     = {10.1142/S0219720008003576}
}

@inproceedings{LiWXY2008,
  author    = {Cindy Y. Li and Prudence W. H. Wong and
               Qin Xin and Fencol C. C. Yung},
  title     = {Approximating Border Length for {DNA}
               Microarray Synthesis},
  booktitle = {Theory and Applications of Models of Computation
               (TAMC 2008)},
  series    = {Lecture Notes in Computer Science},
  volume    = {4978},
  pages     = {410--422},
  publisher = {Springer},
  year      = {2008},
  doi       = {10.1007/978-3-540-79228-4_36}
}

\appendix
\section{Additional Proofs for Hardness Results}\label{app:hardness}

\subsection{Common-prefix cancellation}\label{app:lem-prefix}

\begin{proof}[Proof of Lemma~\ref{lem:prefix}]
Prepending a universally active $c$-column shows $\OPT(cS)\le\OPT(S)$. Conversely, in any alignment of $cS$, erase the first deposited $c$ from every vertex $v$, at time $t_v$. This leaves an alignment of $S$. On an edge $uv$ with $t_u<t_v$, deleting the earlier deposit decreases the border by one, while deleting the later increases it by at most one; if $t_u=t_v$, their contribution is unchanged. Thus total border does not increase. Iterate over the letters of a common prefix. Reversal proves the suffix claim.
\end{proof}

\subsection{Grid coordinate rigidity}\label{app:lem-rigidity}

\begin{proof}[Proof of Lemma~\ref{lem:rigidity}]
Distinct coordinate labels have Manhattan distance at least one, yielding $C(\pi)\ge M$. Equality means that the inverse labeling maps every physical grid edge to an edge of the original grid. Distinct physical edges have distinct images under a vertex bijection, and the two graphs have the same number $M$ of edges. Therefore the map is an edge-preserving bijection, hence a graph automorphism. The converse is immediate. Since $C(\pi)$ is integer, failure of equality yields $C(\pi)\ge M+1$.
\end{proof}

\subsection{Binary encoding of Hamming distances}\label{app:lem-binary}

\begin{proof}[Proof of Lemma~\ref{lem:binary}]
Let $H=\operatorname{Ham}(x,y)$ and $Z=(2d-1)T$, the total number of separator zeros. Aligning corresponding separators and short one-runs yields $\lcs(F(x),F(y))\ge Z+3d-H$, so an optimal LCS deletes at most $H\le d$ positions per word. Number zeros by their rank. Corresponding matched zeros differ in rank by at most $d$. Since every long zero-run has length $T>4d$, a central zero from each run survives and matches inside the corresponding run in the other word. These anchors force any matched one to remain in its corresponding short run. The total number of matchable ones is therefore at most $3d-H$, and at most $Z$ zeros match. Hence $\lcs(F(x),F(y))=Z+3d-H$ and the insertion--deletion distance equals $2H$.
\end{proof}

\section{Additional Proofs for Approximation Algorithms}\label{app:approx}

\subsection{Euler ordering for a tree metric}\label{app:lem-euler}

\begin{proof}[Proof of Lemma~\ref{lem:euler}]
We use the grid-cut argument of Li et al.~\cite{Li2018}. Removing an edge $e$ of the tree partitions its leaves into $A_e$ and $B_e$. Let $s_e=\min\{|A_e|,|B_e|\}$ and let $w_e$ be the weight of $e$. In any grid placement, the contribution of this tree edge is $w_e$ times the number of grid edges joining the two parts. A set of $s\le n/2$ cells of a square grid has at least $c\sqrt{s}$ boundary edges for some constant $c>0$. Thus every placement pays at least $cw_e\sqrt{s_e}$ for this tree edge.

In a depth-first order, the leaves on one side of $e$ form a consecutive interval in the cyclic order. Place the leaves in this order, row by row. Only a constant number of horizontal grid edges cross the interval boundary. At most $O(\min\{m,s_e\})$ vertical edges cross it: only positions within distance $m$ of the ends of the interval can contribute, and the smaller part contains just $s_e$ leaves. The cost due to $e$ is therefore $O(w_e\min\{m,s_e\})$.

The ratio of this contribution to its lower bound is at most
$O(\min\{m/\sqrt{s_e},\sqrt{s_e}\})=O(\sqrt m)=O(n^{1/4})$.
Summing over all tree edges proves the result.
\end{proof}

\subsection{Equal-length probes}\label{app:prop-shortsync}

\begin{proposition}\label{prop:shortsync}
If all probes have length $\ell$, there is also an $O(\ell n^{1/4})$ approximation for square-grid \BMP{} (\cref{prob:bmp}), using the known synchronous placement guarantee of~\cite{Li2018}.
\end{proposition}
\begin{proof}
In a \emph{synchronous} synthesis, probes deposit the letters in the same string position together; different letters in that position require two separate steps. Hence, for equal-length words $x,y$, the synchronous per-edge cost is $2\operatorname{Ham}(x,y)$, while the asynchronous pairwise lower bound is $d(x,y)=2(\ell-\lcs(x,y))$. If $x\ne y$, $\ell-\lcs(x,y)\ge1$ and $\operatorname{Ham}(x,y)\le\ell$, giving $2\operatorname{Ham}(x,y)\le\ell d(x,y)$; equality of words causes zero cost for both models. Thus at an arbitrary fixed layout, synchronous cost is at most $\ell$ times the pairwise insertion--deletion proxy. Apply the $O(n^{1/4})$ synchronous placement algorithm and its exact synchronous embedding, and compare its layout proxy to that of the optimal asynchronous solution as in the proof of Theorem~\ref{thm:mainapprox}.
\end{proof}

\subsection{Three probe types}\label{app:thm-threetypes}

\begin{proof}[Proof of Theorem~\ref{thm:threetypes}]
By \cref{thm:fewtypes}, at most two types can be aligned optimally. Suppose there are exactly three types $x,y,z$. Write $a=d(x,y)$, $b=d(y,z)$, and $c=d(z,x)$. Let $W_a,W_b,W_c$ be the numbers of grid edges joining the respective pairs of types. Every solution has cost at least $L=W_aa+W_bb+W_cc$.
Consider each of the three trees on $x,y,z$. Align its two edges optimally using \cref{lem:treealign} and give all copies of the same type the same embedding. For instance, if we omit the pair $x,y$, the tree contains edges of lengths $b,c$. Its alignment costs at most $L+W_a(b+c-a)$ by the Hamming triangle inequality. The other omitted pairs give analogous bounds.

Put $e_a=b+c-a$, $e_b=c+a-b$, and $e_c=a+b-c$. These numbers are nonnegative by the triangle inequality. If one is zero, omitting its pair gives cost $L$, which is optimal. Otherwise, let $S=a+b+c$. Since $e_a+e_b+e_c=S$, Cauchy--Schwarz gives $\sum_i1/e_i\ge9/S$. Also $a=(S-e_a)/2$, and similarly for $b,c$. Thus $a/e_a+b/e_b+c/e_c$ is at least $(S/2)(9/S)-3/2=3$.
Let $t=\min\{W_ae_a,W_be_b,W_ce_c\}$. Since $L\ge t(a/e_a+b/e_b+c/e_c)$, we have $t\le L/3$. The best of the three tree alignments therefore costs at most $L+t\le4L/3\le4\OPT/3$. All three alignments are computed in polynomial time.
\end{proof}

\subsection{Few distinct probe types}\label{app:thm-fewtypes}

\begin{proof}[Proof of Theorem~\ref{thm:fewtypes}]
Contract identical \emph{types for constructing the alignment only}, retaining the original interaction-edge multiplicities as nonnegative weights $W_{xy}$. Sample an FRT hierarchy on the $k$ distinct words with metric $d$. By Lemma~\ref{lem:hst} there exists an actual type spanning tree $H$ whose path metric satisfies $d_H(x,y)\le4d_T(x,y)$. Align the types along $H$ by Lemma~\ref{lem:treealign} and copy each resulting padded row to all occurrences of that type. Its border cost is at most $\sum_{x<y}W_{xy}d_H(x,y)$. Taking expectations yields $O(\log k)\sum_{x<y}W_{xy}d(x,y)\le O(\log k)\OPT$. When $k=1$, one uniform global alignment has zero border. When $k=2$, align the two types to an LCS and copy their rows to every occurrence: unlike-type edges attain their pairwise lower bound and like-type edges cost zero, so this is optimal.
\end{proof}

\subsection{Length-two probes in one dimension}\label{app:thm-eightseven}

\begin{proof}[Proof of Theorem~\ref{thm:eightseven}]
By the metric triangle inequality, there is an optimal ordering in which copies of each distinct word appear consecutively. Let $k$ denote the number of types. For distinct length-two probes, the insertion--deletion distance is $2$ if their LCS has length one and $4$ if their LCS has length zero. Construct a graph $J$ on the $k$ types joining precisely pairs of distance $2$. If a Hamiltonian ordering uses $g$ edges of $J$, its exact BMP cost is $4(k-1)-2g$. Maximizing $g$ over orderings is exactly the maximum number $g^*$ of edges in a vertex-disjoint path cover of $J$: edges selected from a Hamiltonian ordering form disjoint paths after discarding non-$J$ transitions, and conversely paths of any disjoint path cover can be concatenated with arbitrary interpath transitions. The maximum path-cover algorithm of Berman and Karpinski~\cite{BermanKarpinski2006} returns a path cover with at least $6g^*/7$ edges. Concatenate its paths, reinsert duplicate probes in contiguous zero-cost blocks, and use exact one-dimensional fixed-order alignment. The output cost is at most $4(k-1)-12g^*/7$. Dividing by $4(k-1)-2g^*$, and using $g^*\le k-1$, shows that the ratio is at most $8/7$. The case $k\le1$ is trivial.
\end{proof}

\subsection{Narrow rectangular grids}\label{app:thm-thin}

When the fixed-placement grid has only $r$ rows, a family of spanning trees yields an approximation ratio depending on $r$, not on the number of columns.

\begin{theorem}\label{thm:thin}
For fixed-placement \PBMP{} (\cref{prob:pbmp}) on an $r\times m$ grid with $r\le m$, a deterministic polynomial-time $(r+1)$ approximation exists for arbitrary alphabets and probe lengths.
\end{theorem}
\begin{proof}
For each spine row $a\in[r]$, form a spanning tree $T_a$ containing all vertical grid edges and only the horizontal edges in row $a$. Align exactly along $T_a$ using \cref{lem:treealign} and return the cheapest solution. For a non-tree horizontal edge, bound its mismatch by the tree path descending to row $a$, crossing horizontally, and returning. Averaged over the $r$ choices, each horizontal metric edge has coefficient $1$. A vertical edge between rows $j,j+1$ receives its own coefficient $1$ and at most $4j(r-j)/r\le r$ additional tree-path charges from the horizontal edges. Thus the average cost is at most $(r+1)\sum_{e\in E}d_e\le(r+1)\OPT$, so at least one spine yields the stated ratio.
\end{proof}

\subsection{Grid concentration}\label{app:lem-concentration}

\begin{proof}[Proof of Lemma~\ref{lem:concentration}]
Each exceptional row contains at least one horizontal grid edge of positive total border cost, and each exceptional column contains at least one vertical such edge. These are distinct edges, so there are at most $b$ of either. Every nonexceptional row has the same padded embedding in all its cells, and similarly for nonexceptional columns. If both sorts exist, their pairwise intersections force all the nonexceptional rows and columns to use the same padded word $E^*$. Every cell not simultaneously in an exceptional row and an exceptional column lies in a nonexceptional row or a nonexceptional column and hence has embedding $E^*$. There are at most $b^2$ remaining intersections.
\end{proof}

\subsection{One-dimensional instances with few distinct strings}\label{app:thm-1dk}

\begin{proof}[Proof of Theorem~\ref{thm:1dk}]
For any input ordering, erase all but the first occurrence of each repeated string. The resulting subsequence of types has no higher path cost because $d$ is a metric: shortcutting successive edges cannot increase the total. Conversely, replace each retained type by a contiguous block of all its copies at zero internal distance. Thus there is an optimum with one block per distinct type, and it suffices to find a shortest Hamiltonian path in the complete $d$-weighted metric on those types. Compute all $d$ values by LCS and use Held--Karp dynamic programming with states $(X,v)$, representing the cheapest path visiting exactly type subset $X$ and ending at $v$, with transition $DP[X\cup\{v\},v]=\min_{u\in X}\{DP[X,u]+d(u,v)\}$. The optimum over final endpoints and the associated exact fixed-order tree alignment give the claimed complexity. The bound also implies FPT for the combined alphabet-size-plus-length parameter because $k\le\sum_{i=0}^L|\Sigma|^i$.
\end{proof}

\section{A Polynomial-Time Case}\label{app:exact}

\subsection{Two distinct probe types}\label{app:thm-twotypes}

When only two strings occur, one common LCS alignment realizes the lower bound on every grid edge. The remaining task is to minimize the boundary between the two probe types.

\begin{theorem}\label{thm:twotypes}
If an $h\times h$ free-placement \BMP{} instance (\cref{prob:bmp}) has $A$ occurrences of $x$ and $h^2-A$ occurrences of $y$, then $\OPT=d(x,y)\min_{|X|=A}|\delta_{G_h}(X)|$. It is exactly computable in $O(h^5+|x||y|)$ time.
\end{theorem}
\begin{proof}
Fix a placement and let $X$ be the cells containing $x$, so $|X|=A$. Every edge crossing from $X$ to its complement joins $x$ to $y$ and costs at least $d(x,y)$. One LCS alignment of $x$ and $y$, copied to all occurrences of the corresponding type, achieves this lower bound on every edge. Hence the optimum equals $d(x,y)$ times the smallest cut $|\delta_{G_h}(X)|$ among sets of $A$ cells.

We show that a minimum cut has a simple shape. Let $a_i$ be the number of cells of $X$ in row $i$. Move these cells to the leftmost $a_i$ positions of their row. This cannot increase the horizontal boundary, and the boundary between rows $i$ and $i+1$ becomes $|a_i-a_{i+1}|$, the smallest possible for those row counts. Next sort the row lengths in nonincreasing order. Their total variation cannot increase. We obtain an optimal \emph{Ferrers shape} with $h\ge a_1\ge\cdots\ge a_h\ge0$ and $\sum_i a_i=A$. Its boundary is $\sum_{i=1}^h \mathbf 1_{0<a_i<h}+a_1-a_h$.
A dynamic program records the row index $i$, the number of chosen cells $z$, and the last row length $u$. Extending by a row of length $v\le u$ adds $v$ cells and increases the boundary by $u-v+\mathbf1_{0<v<h}$. There are $O(h^4)$ states and at most $h$ choices per transition. We compute $d(x,y)$ by the LCS algorithm, giving the claimed total running time.
\end{proof}
\end{document}